# Computational biology approach to uncover hepatitis C virus helicase operation

Holger Flechsig

Department of Physical Chemistry, Fritz Haber Institute of the Max Planck Society, Faradayweg 4-6, 14195 Berlin, Germany

**Abstract**

Hepatitis C virus helicase is a molecular motor that splits nucleic acid duplex structures during viral replication, therefore representing a promising target for antiviral treatment. Hence, a detailed understanding of the mechanism by which it operates would facilitate the development of efficient drug-assisted therapies aiming to inhibit helicase activity. Despite extensive investigations performed in the past, a thorough understanding of the activity of this important protein was lacking since the underlying internal conformational motions could not be resolved. Here we review investigations that have been previously performed by us for HCV helicase. Using methods of structure-based computational modelling it became possible to follow entire operation cycles of this motor protein in structurally resolved simulations and uncover the mechanism by which it moves along the nucleic acid and accomplishes strand separation. We also discuss observations from that study in the light of recent experimental studies that confirm our findings.

# 1. Introduction

Hepatitis C is a liver disease caused by infection with the hepatitis C virus (HCV). According to the world health organization (WHO), about 150 million people are chronically infected by HCV with 3 to 4 million additional infections every year. About 75-85% of newly infected individuals develop chronic disease with long-term complications including liver cirrhosis and liver cancer. Since to date no vaccine is available and more than 350,000 patients die yearly in consequence of hepatitis C disease, the virus poses a major health concern.

HCV belongs to the family of *Flaviviridae* and has a positive sense single-stranded RNA genome that consists of a single open reading frame, which is translated into a single protein of approx. 3,000 amino acids. This polyprotein is further cleaved by viral and cellular proteases into three structural and seven non-structural proteins, which organize replication and packaging of the viral genome. A key role in the multiplication process of the virus is occupied by non-structural proteins NS5A and NS3, which function as molecular motors directly interacting with RNA. NS5A is a polymerase that synthesizes new viral RNA while NS3 contains a large portion that operates as a helicase, capable of translocating along single nucleic acids strands and separating their complementary strands or remove proteins bound to them[1]. Due to their eminent role in the replication machinery of HCV, these proteins have attracted much attention as possible targets for anti-HCV drugs and thus towards the development of effective treatments[2-4].

## 1.1 Hepatitis C virus helicase

Helicases are motor proteins that are powered by adenosine-triphosphate (ATP) molecules and ubiquitous in virtually all processes that involve manipulation of DNA and RNA[5]. They are capable of displacing proteins bound to the nucleic acids, removing secondary structures and, most prominently, separating their duplex structures into single-stranded components[6,7]. Hence, they play a central role in replication, recombination and repair of nucleic acid substrates. Likewise generally in molecular motors, the operation of helicase proteins relies on cyclic internal conformational motions driven by binding of an ATP molecule, its hydrolysis into adenosine-diphosphate (ADP) and a phosphate $P_i$, and the dissociation of the chemical products[8,9]. These internal structural changes are used to generate forces on the nucleic acid strand and allow the helicase to translocate along it and eventually implement its particular function.

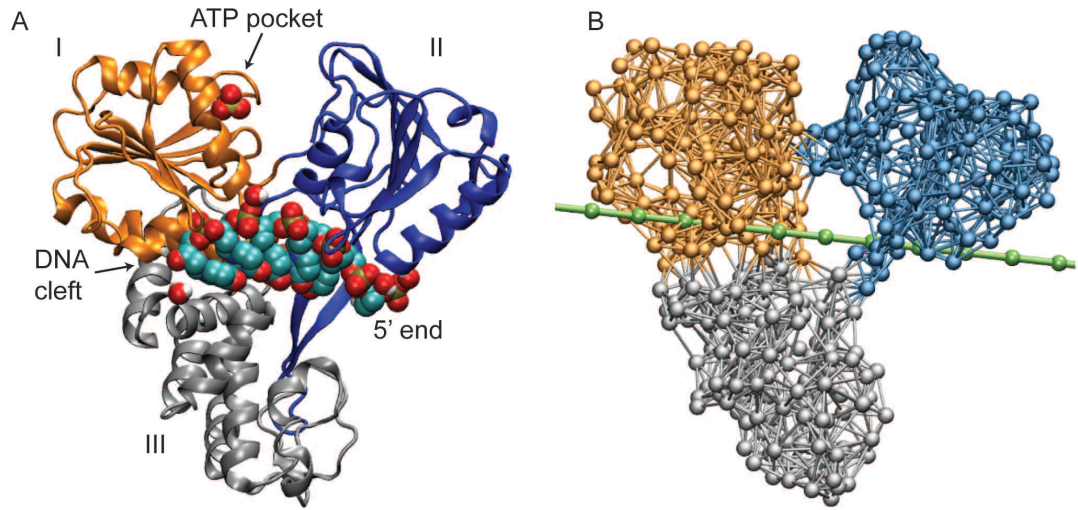

**Fig. 1: Molecular architecture of HCV helicase. (A)** Structure of HCV helicase in ribbon representation (PDB code 1A1V). Motor domains I and II are coloured golden and blue, respectively. Domain III is shown in grey. A sulfate ion, shown as beads, is located at the surface of domain I in the proximity of the ATP binding pocket. The short DNA fragment shown in bead representation occupies a large cleft separating the motor domains from the third domain. **(B)** Network representation of HCV helicase. The chain of beads representing a single DNA strand is shown in green.

The NS3 helicase was the first HCV protein that could be successfully crystallized and to date represents the probably best-characterized helicase molecule. The first structures of this protein were reported in 1997 by Yao et al. and, bound to a short fragment of DNA oligonucleotides, in 1998 by Kim et al[10,11]. Figure 1 shows the molecular architecture of HCV helicase. The protein consists of three domains that are arranged in a Y-shaped form. Domains I and II are structurally related and referred to as the helicase motor domains. At the open interface between these domains, ATP molecules can arrive and bind to a pocket that is located close to the surface of domain I. The motor domains are separated from the third helicase domain by a large deep cleft in which a single nucleic acid strand can be bound and hold.

In the past the HCV helicase protein has been the subject of numerous investigations, aiming to explore its functional properties and understand the mechanism by which it operates. Despite its clinical relevance it has thus become also a prototype helicase representing a broad class of structurally and functionally related proteins[12]. Already the first crystallographic studies of HCV helicase have indicated the mobility of motor domain II, which is flexibly linked to motor domain I, and discussed its possible functional role. Based on the early studies, time-resolved single molecule experiments have been designed and important aspects of HCV helicase operation were investigated in optical tweezer setups and using fluorescence analysis[13,14]. These experiments have elucidated the ATP-dependent unwinding activity of the helicase on DNA and RNA duplex structures and greatly contributed to the understanding of helicase function. Although models for the operation of HCV helicase have been established and theoretical investigations were performed additionally[6,11,15], the internal conformational motions related to interactions with ATP molecules in this protein could not be sufficiently resolved and thus the particular mechanism by which HCV helicase moves along nucleic acid strands and performs strand separation remained elusive hitherto. In this situation, modelling studies that are based on present structures of HCV helicase and employ methods from computational biology may play an important role.

In this article we review investigations that have been previously performed by us for HCV helicase. We shall focus here on the conceptual aspects of our analysis and present our findings in a rather general framework, while a more profound description including details can be gathered from our original publication[16]. Moreover, we will discuss our findings in the light of novel crystal structures of HCV helicase complexed with ATP-mimicking nucleotides and DNA and RNA.

## 1.2 Computational Biology Methods

The function of any protein relies on its intrinsic structural flexibility and is caused by internal conformational motions. Since these motions can be hardly resolved to a sufficient extend by experimental methods and, furthermore, available structural data provides only static snapshots, capturing a protein at a few steps along its operation, our knowledge about the operation principles of proteins is often limited. On the other side, the tremendous increase of computational power that has become available over the last decades have enabled the investigation of protein dynamics in computer simulations and opened promising new possibilities in protein research employing numerical modelling.

Widespread approaches that have become a standard tool of computer-assisted molecular biology are molecular dynamics simulations[17-19]. These descriptions take into account all atoms of a protein and allow to model complicated interactions between them by using multi-parameter force-fields. Molecular dynamics simulations allow to investigate dynamical processes in proteins at high resolution. Their application, however, is limited by the size of the protein and the timescales of internal motions. While usually these methods are capable of resolving rapid processes with motions on timescales up to nanoseconds, they fail to cover the biologically relevant slow conformational motions associated with the ATP-related turnover cycles in protein machines and molecular motors. To follow these large-amplitude motions with typical timescales of tens of milliseconds in molecular dynamics simulations would require a massive computational expense, which is beyond what can be reached nowadays, even with modern computer architectures.

To overcome these difficulties, coarse-grained models of protein dynamics have been developed and became increasingly popular in the past[20]. Being based on rigorous approximations, they allow a straightforward numerical implementation and can drastically reduce the computational burden. In such models, the structure of a protein is simplified and the complicated intra-molecular interactions are replaced by effective potentials. A coarse-grained approach commonly used to investigate dynamics of protein machines is the elastic network model[21,22]. Within this description, entire amino acid residues of a protein structure are typically replaced by single identical beads and interactions between them are mediated by empiric harmonic potentials that depend only on the distance between any two beads. Within the framework of this model, a protein is therefore viewed as an elastic object, i.e. a network of beads connected by deformable elastic springs. Despite the coarse-grained nature of elastic network descriptions and the gross simplification made, these models have been proved powerful in describing ATP-induced slow motions in molecular machines and advanced the understanding of important aspects of their collective dynamics[23-28].

In the first part of this review we will explain how an elastic-network description of HCV helicase that incorporated interactions with ATP-ligands was established and individual operation cycles were followed in computer experiments. We will then show how the dynamical modelling was extended to include also interactions with nucleic acid strands, which allowed us to study how internal conformational motions are coupled to progressive helicase locomotion and follow the mechanism of strand separation.

## 2. ATP-induced operation cycles of HCV helicase

We have used the nucleotide-free structure determined by Yao et al. in 1997[10] (Protein Data Bank (PDB) code 1HEI) to construct the elastic network of HCV helicase. The network (shown in Figure 1B) consisted of 443 identical beads, which represented entire amino acid residues, and elastic springs connecting any two beads if their spatial distance was below a prescribed interaction radius of 8 Angstroms. A spring connecting two beads was assigned the natural length that corresponded exactly to the distance between the two beads as extracted from the PDB file. Therefore, by construction, the initial network represented a steady conformation of the protein. When, however, interactions between the network and ligands are taken into account, network dynamics can be induced and related conformational motions analyzed. The internal motions of proteins in solution are dominated by friction and internal forces are negligible already on time-scales slower than tens of picoseconds. The slow ATP-related conformational motions in protein machines and molecular motors are certainly overdamped. Therefore, within the elastic network model, slow conformational motions of a protein are described as relaxation motions of its corresponding elastic network. If for simplicity thermal fluctuations and hydrodynamical interactions are neglected, the dynamics of the network can be described by equations of motion that correspond to Newton mechanics (see [16]). This set of equations is numerically integrated in the computer simulations to obtain the position of all network beads at every instant of time. In this way structural changes of the protein network can be conveniently traced.

To follow ATP-induced dynamics in HCV helicase, we have established a ligand-elastic-network model that incorporated binding of ATP, its hydrolysis, and the dissociation of the product in an approximate fashion. ATP itself is a relatively complex molecule and its interactions with the protein rely on complicated processes and involve chemical details that cannot be accounted for in our description. To remain consistent with the approximate nature of the model, we have described an ATP molecule by a single ligand bead (the substrate ligand) and modelled its binding process to the ATP-pocket by allowing it to physically interact with network beads present there. Specifically, upon binding the substrate ligand could create spring connections to four selected beads inside the binding pocket. These particular interaction beads corresponded to amino acid residues that belong to sequence motifs, which have been shown to be conserved among all helicases and are crucially involved in interactions with ATP. The additional ligand springs were initially stretched so that initiated by the binding process, attractive forces between the ligand and the four pocket beads were generated and network deformations became induced. First localized in the binding pocket they gradually spread and resulted in conformational motions of the entire protein network until eventually a steady network state that corresponded to the network-ligand complex was reached. In this conformation, to roughly emulate the hydrolysis of ATP, we assumed that the substrate ligand became converted into the product ligand, which represented the chemical products ADP and $P_i$ being ready to dissociate from the protein. Therefore, we assumed that the product ligand, in contrast to the substrate ligand, could no longer interact with the ligand pocket and thus removed the four ligand springs. Since then, the ligand-free network found itself in a conformation that was different from the native one, with the network springs being deformed, conformational motions bringing it back to its initial state occurred. The reset ligand-free network could then in principle bind another substrate ligand and the next ATP-induced operation cycle would be initiated.

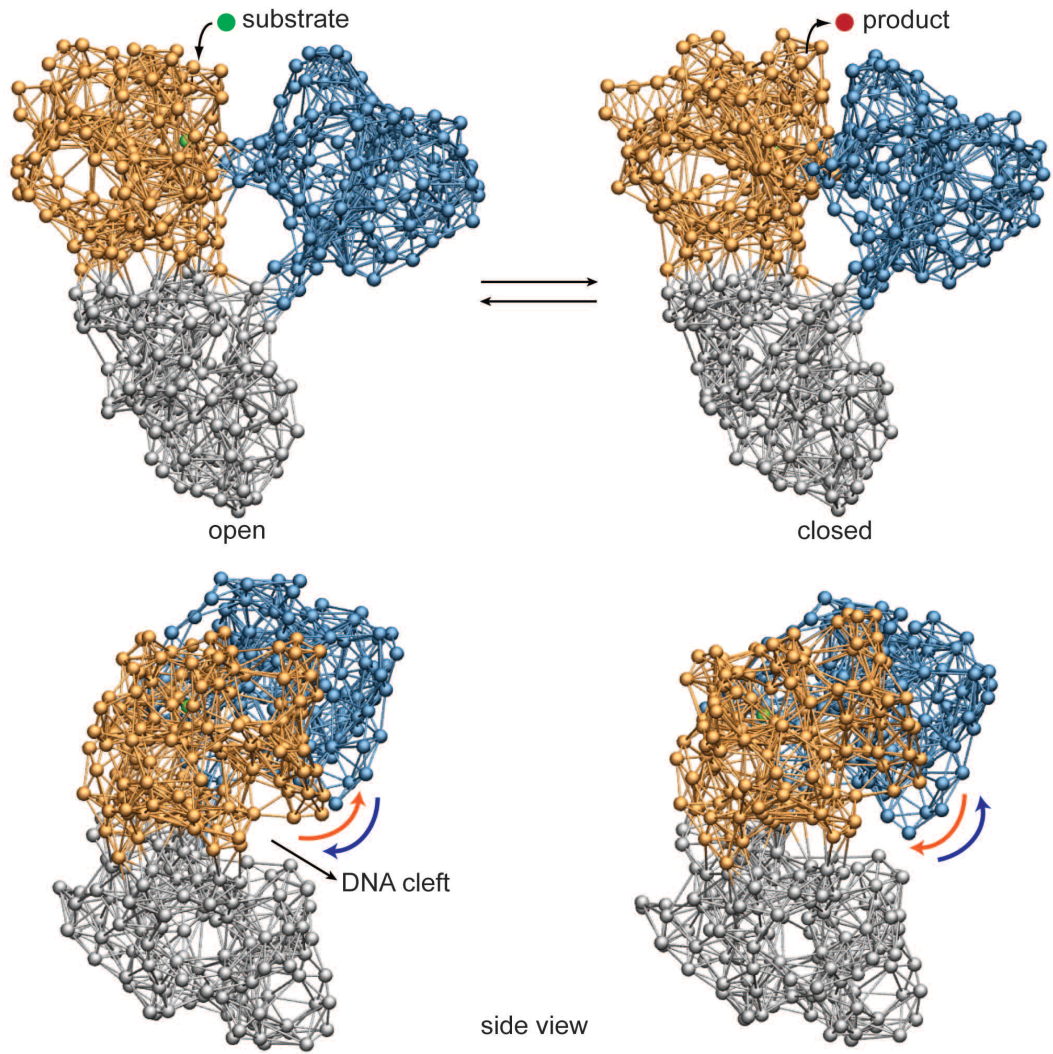

**Fig. 2: Cyclic ligand-induced conformational motions in HCV helicase.** Top row: binding of the substrate ligand led to switching from open to closed conformation. Release of the product ligand induced motions back to the open form. Bottom row: motions of the motor domains relative to the nucleic acid binding cleft that are critical for grip control are indicated by arrows.

The cyclic ligand-induced conformational changes in HCV helicase as obtained from our simulations are depicted in Figure 2. We observed that subsequent to binding of the substrate ligand, pronounced relative motions of the two motor domains were induced, bringing the HCV helicase from an open-shape conformation in which these domains were well-separated into a closed-shape form where they were found in tight contact. In this conformation of the network-ligand complex, the substrate ligand was converted into the product ligand, which was then removed from the network. Under subsequent conformational motions, bringing the helicase network back to its initial configuration, the motor domains moved away from each other and restored the open-shape form.

In addition to these large-scale domain rearrangements, we recognized another component of the same dynamics that was more subtle and consisted in motions of the domains with respect to the cleft that holds the nucleic acid strand. Such changes become apparent in the side view perspective of the protein (see Fig. 2): Under binding of the ligand, motor domain I moved upwards and therefore broadened the cleft at one side, whereas motor domain II moved downwards narrowing the cleft at the opposite side. Along the back-motion of the network, after removal of the product ligand, the situation was reversed. Domain I moved down again whereas domain II was lifted away from the nucleic acid cleft.

Our findings suggest that by performing these motions relative to the cleft that holds the nucleic acid strand, each of the motor domains is capable of controlling its grip on the strand. Consequently, by weakening or tightening the grip depending on the presence of the substrate ligand, the large-amplitude opening and closing motions of the motor domains can be translated into steady translocation along the single nucleic acid strand by means of a ratchet mechanism. To demonstrate the mode by which HCV helicase is able to move along nucleic acid strands, we have performed simulations in which an approximate description of a DNA strand was included and interactions between the protein and DNA were accounted for in a simplified way.

### 3. Ratcheting inchworm translocation

We have extended our dynamical description to incorporate interactions between the HCV helicase and a single DNA strand. While apparently detailed models of DNA and its dynamical properties are available, it was more important for our purposes, to employ a proper mechanical description of this molecule. To remain consistent with the coarse-grained modelling implemented so far, a single DNA strand was described as a chain of identical beads, which were consecutively linked by elastic springs (see Fig. 1B). The chain, which was representing the sugar-phosphate backbone of DNA, was placed roughly in the center of the nucleic acid binding cleft and was allowed to undergo deformations such as stretching and bending when interacting with the helicase.

Regarding the interactions between HCV helicase and the nucleic acid, it is known that instead of being distributed over the entire cleft, major contacts to the sugar-phosphate backbone of DNA are established to specific conserved amino acid residues of the protein[11]. These are Thr269 from motor domain I and Thr411 from motor domain II. In our description we have restricted helicase-DNA interactions to those two key residues and modelled them only roughly via additional elastic links that were established when a motor domain moved towards the DNA strand and removed when a motor domain was lifted from the strand. As described in the previous section, upon binding of the substrate ligand to the helicase network motor domain II moved down and therefore tightened its grip on the DNA strand (hand-on state) whereas at the same time motor domain I moved away from the strand and weakened its grip (hand-off state). Therefore, when the substrate ligand was bound, we have created a stiff bond between the Thr411-bead from motor domain II and the nearest bead in the DNA chain, while a bond connecting the Thr269-bead from motor domain I to the chain was absent. Then, along the formation of the network-ligand complex, the large-amplitude slow conformational changes occurred under which motor domain I moved towards domain II until they were in tight contact. At this stage, when the ligand after its conversion from substrate into product was removed, the grip of the motor domains on the DNA was reversed. Consequently, the bond between motor domain II and the DNA was removed (hand-off) and at the same time a bond between the Thr269-bead from motor domain I and the nearest bead in the DNA was established (hand-on). Then, during the ligand-free part of the operation cycle, the motor domains slowly moved away from each other, this time with motor domain I holding the DNA chain. After one ligand-induced operation cycle, the helicase is found again in the initial open-shape conformation with the motor domains being separated, but became effectively transported along the DNA - as we found - by one single base in the chain. The translocation mechanism is illustrated in Figure 3 where snapshots from the simulation are depicted.

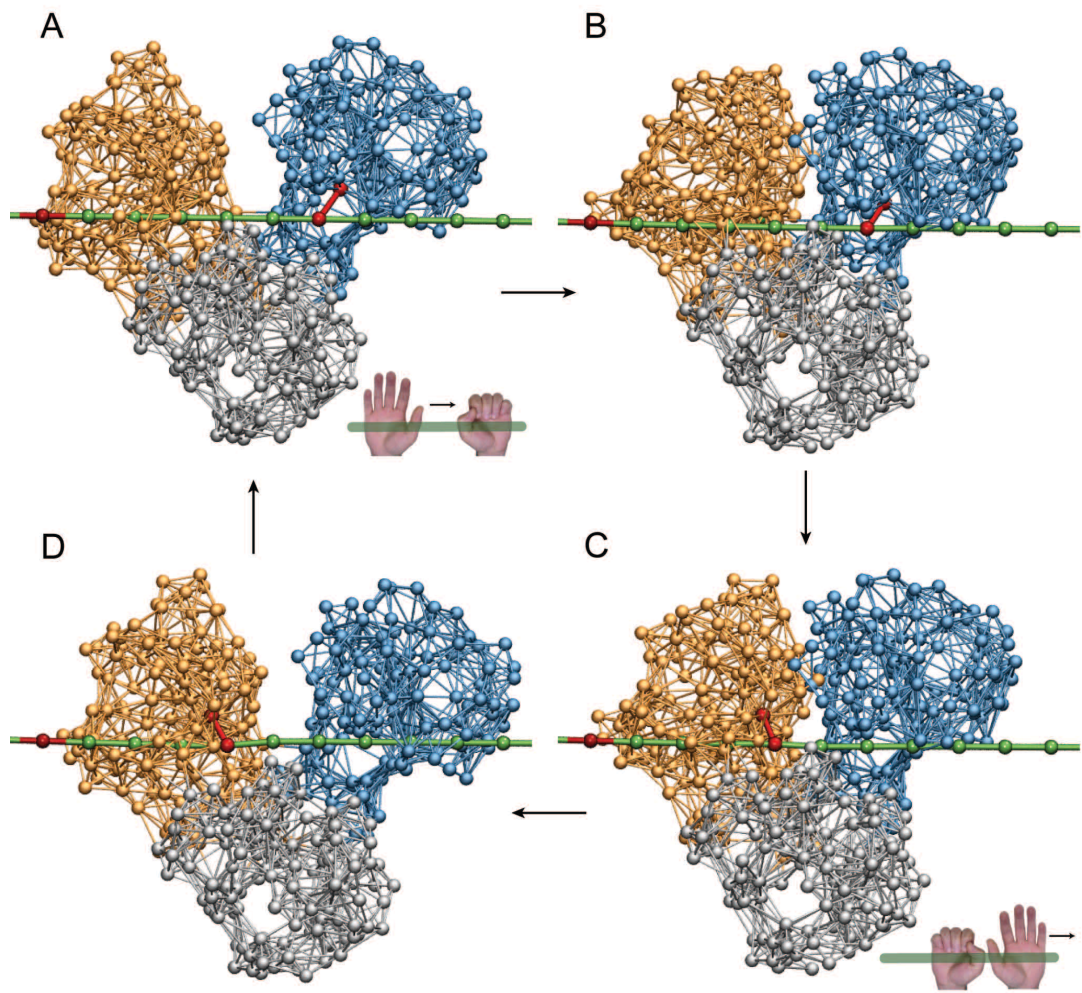

**Fig. 3: Ratcheting inchworm translocation.** Four consequent snapshots (A-D) within a single ligand-induced operation cycle are shown. Red bonds indicate connections between a motor domain and the DNA chain. After one cycle, the helicase is found in a similar open-shape conformation, but is propagated by one base along the DNA strand (compare (A) and (D)).

With the described model extensions we could successfully uncover, in a structurally resolved way, the mechanism by which HCV helicase is able to move along a single DNA strand. The style of locomotion along DNA used by HCV helicase is reminiscent of the mechanism by which worms deform themselves to crawl along trees. The ATP-dependent opening and closing motions of the two motor domains are converted into directed base-by-base translocation along the DNA strand by an integrated ratcheting mechanism of these domains, i.e. the alternating gripping (hand-on hand-off style) on the DNA. Because of that analogy, the described mechanism of propagation along DNA employed by HCV helicase is referred to as the ratcheting inchworm translocation.

## 4. Strand-separation by HCV helicase

In the performed computer experiments, we finally aimed to demonstrate how steady translocation of HCV helicase along the single nucleic acid strand is coupled to the separation of the second complementary strand. Therefore, we have set up a model of duplex DNA that consisted of two opposite single chains that were bridged by additional links. These connections were intended to effectively mimic the attractive interactions between complementary bases of the opposing single chains that stabilize the duplex form of DNA. On the other side, however, these bridge-links could also be broken when the two strands become separated and interactions between them gradually vanished. The duplex DNA was positioned in such a way that the upper strand was again centered in its putative binding cleft and the complementary strand was located below the motor domains of the protein (see Fig. 4).

While the two motor domains of HCV helicase serve as essential modules, whose opening and closing motions make inchworm translocation possible, the role of the third protein domain is less clear. As revealed from our previous simulations, it is rigidly connected to motor domain I and therefore has to follow its motion. Hence, during progressive translocation of the motor domains the third domain will always be dragged by domain I. This fact may suggest an active role of domain III during the process of duplex separation. Although little is known about interactions between HCV helicase and the duplex DNA, we can, based on the previously proposed mechanism[11], assume the following mechanical scenario of action. Under steady translocation of the two motor domains along the upper DNA strand, the third helicase domain encounters the free end of the complementary strand at the DNA fork and interactions between them set in. Particularly, repulsive forces shall appear that tend to push apart the lower DNA strand and prevent it from penetrating into the protein structure where it would cause partial unfolding. In the simulation we have therefore assumed that the lower DNA strand was always subject to an overall repulsive force that was generated by beads from the third helicase domain and which became strong at short distances.

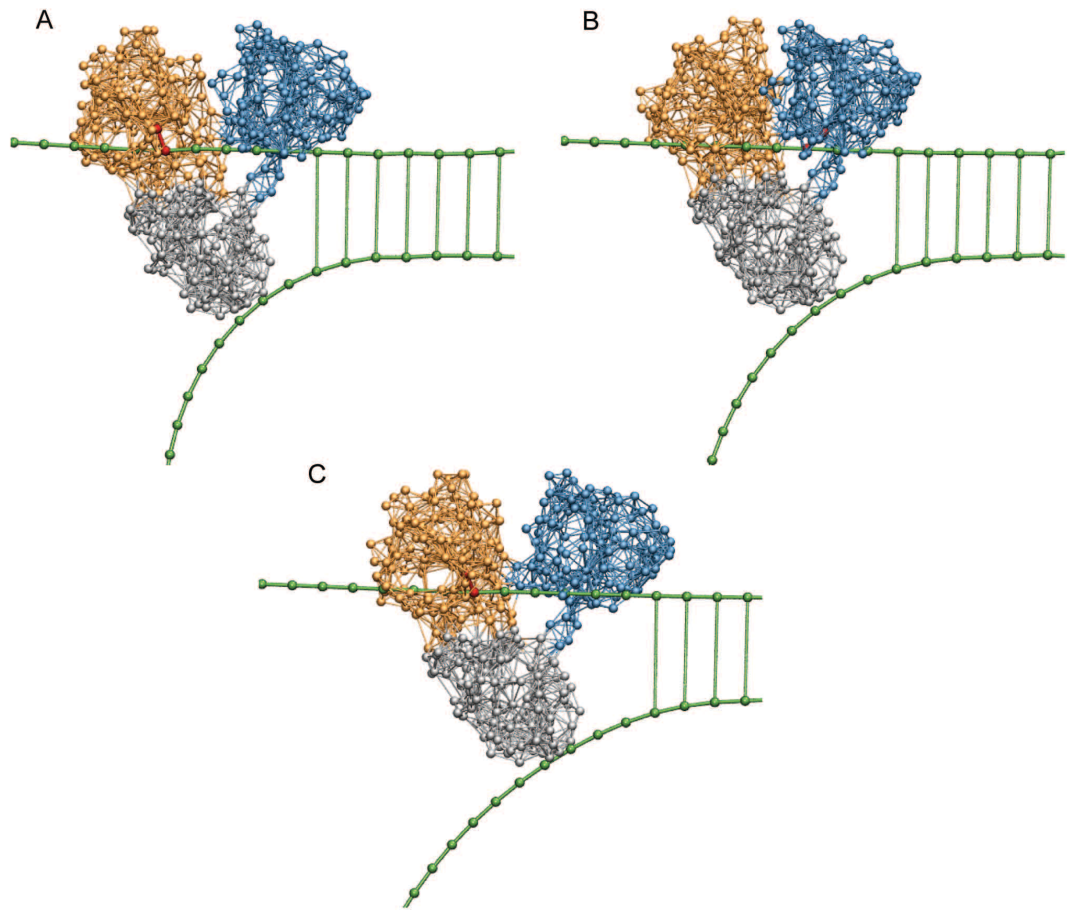

**Fig. 4: Unzipping of duplex DNA by HCV helicase.** Shown are three consequent snapshots (A-C) from a simulation that demonstrates ratcheting inchworm translocation and mechanical strand separation.

With the implemented setup we have traced three ligand-induced operation cycles of HCV helicase within a single simulation and observed how progressive base-by-base inchworm translocation along the upper DNA strand lead to separation of the lower strand. In Figure 4 snapshots taken at different time moments during the simulation are provided. As the two motor domains moved along the upper DNA strand, the third helicase domain was dragged into the free space at the duplex fork. As a result of the repulsive interactions that tried to prevent contact of this domain with the lower strand, forces pushing away that strand were generated. As these forces built up, stress was accumulated in the bridge-links that held together the duplex. Eventually these connections became broken one after another and the duplex DNA was successively unzipped. Our simulation show that the third helicase domain acted like a wedge that, in the course of translocation along the upper strand, was pressed between the two DNA strands and mechanically separated them. We found that HCV helicase had broken three base-pairs after three translocation steps, one base-pair at a time.

## 5. Conclusions and recent experiments

Molecular machines and motors are nanoscale engines that operate by performing cyclic changes of their conformation, powered by the chemical energy that is derived from the hydrolysis of ATP molecules. To observe these conformational motions directly in single-molecule experiments in sufficient detail is however not yet possible due to the limited resolution of nowadays techniques. Therefore important aspects of the activity of motor proteins remained elusive so far. On the other side, the growing capacities of modern computers have facilitated the investigation of protein dynamics in computer simulations.

Here, we present a review of computer-assisted investigations that have been previously performed for the helicase protein of hepatitis C virus (HCV) employing coarse-grained molecular dynamics methods. HCV helicase is a molecular motor that translocates over nucleic acid strands and splits their duplex structure. Since it plays a crucial role in the replication cycle of the virus, HCV helicase presents a major target for drugs and understanding of the functional properties underlying its operation is a key step towards the development of efficient anti-HCV therapies.

Despite a lot of past studies, the mechanism by which HCV helicase moves over nucleic acid and drives strand separation remained speculative. By establishing a structure-based mechanical model of this protein that included interactions with ATP molecules and DNA in an approximate manner, we were able to follow entire operation cycles of HCV helicase in computer experiments. We have found that the ATP-induced dynamics in this protein consisted in well-organized large-amplitude conformational changes and involved concerted relative motions of the two motor domains. These motions are functional and, as identified from our simulations, can lead to progressive base-by-base inchworm translocation along a single DNA strand my means of a ratchet mechanism. Along this process, the third helicase domain acts as a wedge that is pressed between the two DNA strands and actively separates the duplex structure by breaking base-pairs one at a time.

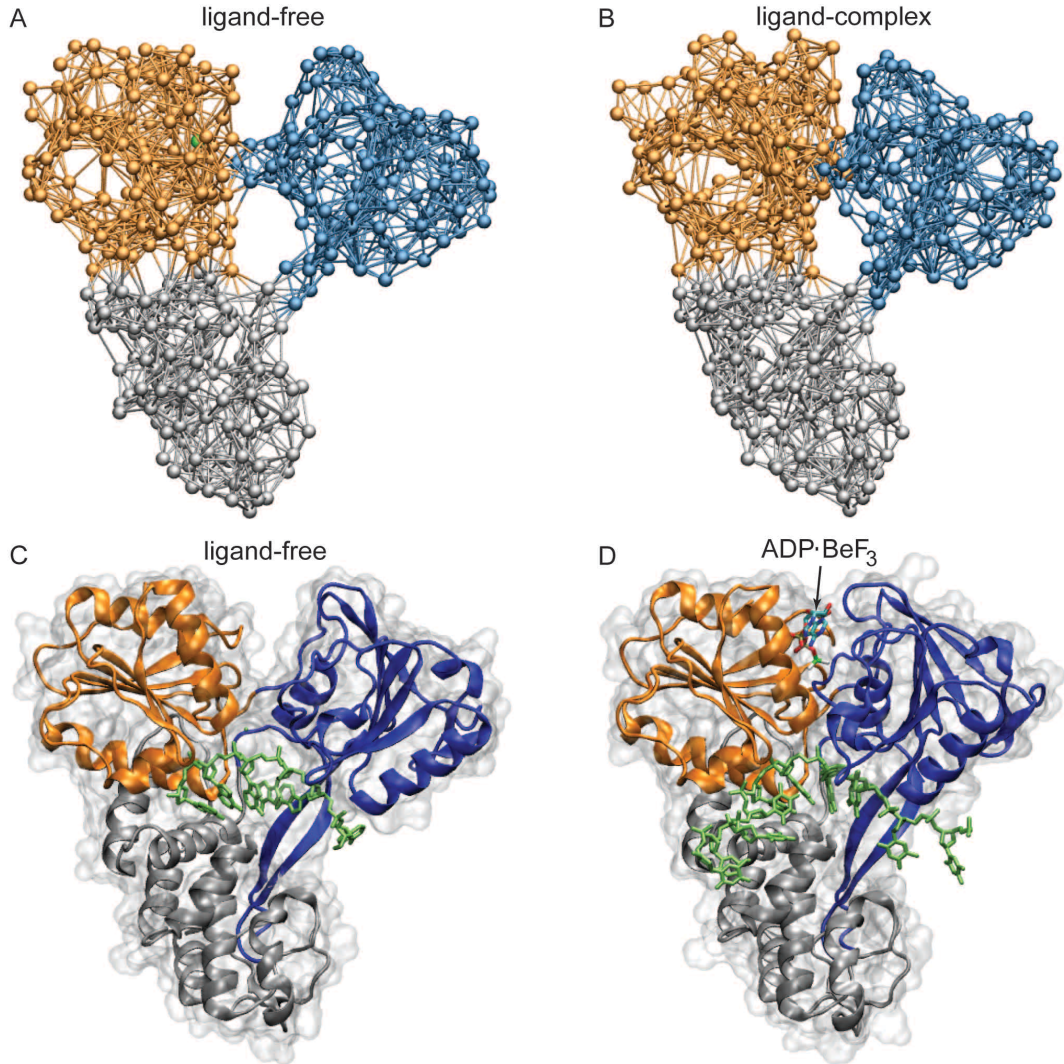

**Fig. 5: Comparison with novel HCV crystal structures.** Structural changes between the ligand-free conformation and the ligand-bound complex as predicted by our simulations (top row) are compared with structures determined in recent experiments (bottom row, PDB codes 3KQH and 3KQU; ADP·BeF$_3$ was used as ATP analog).

The ratcheting inchworm translocation had been previously proposed based on experimental data. Furthermore these results suggested that DNA unwinding occurs in bursts, whereas our findings reveal that base-pairs are broken step by step. After our work was published, Charles M. Rice reported in 2010, in his Inaugural Article in PNAS, novel crystal structures for a set of HCV helicase complexes with ATP-mimics and bound single-stranded DNA[29]. They provide, for the first time, conformational snapshots along the functional cycle of HCV helicase and allow to discuss the molecular basis of its activity. The obtained data shows that upon binding of the ATP-mimicking nucleotide the two motor domains come into tight contact, and, while interactions between domain II and the DNA are maintained, they appear to be decreased between the protein domain I and the 3' end of the DNA. In a transition conformation corresponding to a post-hydrolysis state of the helicase, however, distances to the DNA are changed and new connections between domain I and the DNA strand appear whereas domain II may be able to slide along the strand when returning to the initial open-shape conformation of the helicase. Taken together, these novel crystal structures provide insights into ATP-induced conformational changes in HCV helicase and offer a molecular-level explanation of the ratcheting single-base translocation mechanism based on the observed loss and recovery of interactions between the protein and the single DNA strand. Furthermore, in 2011 crystal structures capturing HCV helicase with its natural substrate RNA became available in its nucleotide-free conformation and bound to a non-hydrolyzable analog[30]. They also evidence large domain motions and the switching from open to closed conformation in response to binding of the ATP-mimicking ligand and point towards a mechanism of unidirectional stepping along RNA by one base-pair per hydrolysis cycle, caused by alternating affinities of the helicase to bind the RNA.

The novel structural data has unravelled important aspects of the molecular mechanisms that underlie the activity of HCV helicase and thereby largely contributed to the understanding of how this important molecular motor performs its operation. In Figure 5, observations from our simulation are compared with the experimentally determined structures. The structural data is in well agreement with the predictions of conformational motions as revealed from our computer experiments, as they confirm both the essential relative motions of the motor domains and the ratchet mechanism employed by the protein to translate them into locomotion along the nucleic acid strand with a step-size of one single base. In addition to the static structures, results from high-resolution optical tweezers experiments, aimed to follow the unwinding of double-stranded RNA by HCV helicase, have been reported in 2011 by the Bustamante lab[31]. They observe an opening of the duplex structure in single base-pair steps, which is consistent with what we have found in our computer experiments. It should be mentioned that recently coarse-grained modelling of HCV helicase has been combined with atomistic simulations[32]. The obtained results are in support of the inchworm model of helicase locomotion and propose key amino acid residues crucial for the translocation machinery.

In summary, using coarse-grained structure-based modelling that included interactions with ATP molecules and DNA in an approximate fashion, we were able to follow entire operation cycles of an important molecular motor in computer experiments for the first time. Our findings are today confirmed by experimental data that has been acquired recently. Generally, our study demonstrates the feasibility of approximate mechanical models in computer experiments of molecular machines and motors. Despite their coarse-grained nature, such dynamical descriptions are capable of explaining functional aspects of these nanoscale engines and therefore help to understand the general principles by which they operate. Although approaches from computational biology have been proven remarkably successful in assisting in protein research, their explanatory power is also limited and apparently they cannot replace the experiments. Recently, large progress has been achieved in single-molecule techniques to investigate protein dynamics. In experiments that are based on the high-speed atomic force microscopy it became possible to observe protein machines as they perform their operation[33,34]. These approaches offer new possibilities towards understanding the important machines that control life at the nanoscale. Researchers may also use such experimental methods to investigate helicase motors in the future.

**References:**

[1] **Frick DN**, Lam AMI. Understanding helicases as a means of virus control. *Curr Pharm Des* 2006; **12**: 1315-1338. doi: 10.2174/138161206776361147.

[2] **De Francesco R**, Migliaccio G. Challenges and successes in developing new therapies for hepatitis C. *Nature* 2005; **436**: 953-960. doi: 10.1038/nature04080.

[3] **Kwong AD**, Govinda Rao B, Jeang KT. Viral and cellular RNA helicases as antiviral targets. *Nat Rev Drug Discovery* 2005; **4**: 845-853. doi: 10.1038/nrd1853.

[4] **Moradpour D**, Penin F, Rice CM. Replication of hepatitis C virus. *Nat Rev Microbiol* 2007; **5**: 453-463. doi: 10.1038/nrmicro1645.

[5] **Caruthers JM**, McKay DB. Helicase structure and mechanism. *Curr Opin Struct Biol* 2002; **12**: 123-133. doi: 10.1016/S0959-440X(02)00298-1.

[6] **Lohman TM**, Bjornson KP. Mechanisms of helicase-catalyzed DNA unwinding. *Annu Rev Biochem* 1996; **65**: 169-214. doi: 10.1146/annurev.bi.65.070196.001125.

[7] **Jankowsky E**, Fairman ME. RNA helicases – one fold for many functions. *Curr Opin Struct Biol* 2007; **17**: 316-324. doi: 10.1016/j.sbi.2007.05.007.

[8] **Spudich JA**. How molecular motors work. *Nature* 1994; **372**: 515-518. doi: 10.1038/372515a0.

[9] **Enemark EJ**, Joshua-Tor L. On helicases and other motor proteins. *Curr Opin Struct Biol* 2008; **18**: 243-257. doi: 10.1016/j.sbi.2008.01.007.

[10] **Yao N**, Hesson T, Cable M, Hong Z, Kwong AD, Le HV, Weber PC. Structure of the hepatitis C virus RNA helicase domain. *Nat Struct Biol* 1997; **4**: 463-467. doi: 10.1038/nsb0697-463.

[11] **Kim JL**, Morgenstern KA, Griffith JP, Dwyer MD, Thomson JA, Murcko MA, Lin C, Caron PR. Hepatitis C virus NS3 RNA helicase domain with a bound oligonucleotide: the crystal structure provides insights into the mode of unwinding. *Structure* 1998; **6**: 89-100. doi: 10.1016/S0969-2126(98)00010-0.

[12] **Frick DN**. The hepatitis C virus NS3 protein: a model RNA helicase and potential drug target. *Curr Issues Mol Biol* 2007; **9**: 1-20

[13] **Dumont S**, Cheng W, Serebrov V, Beran RK, Tinoco I, Pyle AM, Bustamante C. RNA translocation and unwinding mechanism of HCV NS3 helicase and its coordination by ATP. *Nature* 2006; **439**: 105-108. doi: 10.1038/nature04331.

[14] **Myong S**, Bruno MC, Pyle AM, Ha T. Spring-loaded mechanism of DNA unwinding by hepatitis C virus NS3 helicase. *Science* 2007; **317**: 513-516. doi:10.1126/science.1144130.

[15] **Zheng W**, Liao JC, Brooks BR, Doniach S. Toward the mechanism of dynamical couplings and translocation in hepatitis C virus NS3 helicase using elastic network model. *Proteins* 2007; **67**: 886-896. doi: 10.1002/prot.21326.

[16] **Flechsig H**, Mikhailov AS. Tracing entire operation cycles of molecular motor hepatitis C virus helicase in structurally resolved dynamical simulations. *Proc Natl Acad Sci USA* 2010; **107**: 20875-20880. doi: 10.1073/pnas.1014631107.

[17] **McCammon JA**, Gelin BR, Karplus M. Dynamics of folded proteins. *Nature* 1977; **267**: 585-590. doi: 10.1038/267585a0.

[18] **Tuckerman ME**, Martyna GJ. Understanding modern molecular dynamics: techniques and applications. *J Phys Chem B* 2000; **104**: 159-178. doi: 10.1021/jp992433y.

[19] **Karplus M**, McCammon JA. Molecular dynamics simulations of biomolecules. *Nat Struct Biol* 2002; **9**: 646-652. doi:10.1038/nsb0902-646.

[20] **Tozzini V**. Coarse-grained models for proteins. *Curr Opin Struct Biol* 2005; **15**: 144-150. doi: 10.1016/j.sbi.2005.02.005.

[21] **Tirion MM**. Large amplitude elastic motions in proteins from a single-parameter, atomic analysis. *Phys Rev Lett* 1996; **77**: 1905-1908. doi: 10.1103/PhysRevLett.77.1905.

[22] **Bahar I**, Atilgan AR, Erman B. Direct evaluation of thermal fluctuations in proteins using a single-parameter harmonic potential. *Fold Des* 1997; **2**: 173-181. doi: 10.1016/S1359-0278(97)00024-2.

[23] **Hinsen K**. Analysis of domain motions by approximate normal mode calculations. *Proteins* 1998; **33**: 417-429. doi: 10.1002/(SICI)1097-0134(19981115)33:3<417::AID-PROT10>3.0.CO;2-8.

[24] **Bahar I**, Atilgan AR, Demirel MC, Erman B. Vibrational dynamics of folded proteins: significance of slow and fast motions in relation to function and stability. *Phys Rev Lett* 1998; **80**: 2733-2736. doi: 10.1103/PhysRevLett.80.2733.

[25] **Tama F**, Sanejouand YH. Conformational change or proteins arising from normal mode calculations. *Protein Eng* 2001; **14**: 1-6. doi: 10.1093/protein/14.1.1.

[26] **Zheng W**, Doniach S. A comparative study of motor-protein motions by using a simple elastic-network model. *Proc Natl Acad Sci USA* 2003; **100**: 13253-13258. doi: 10.1073/pnas.2235686100.

[27] **Yang L**, Song G, Jernigan RL. How well can we understand large-scale protein motions using normal modes of elastic networks. *Biophys J* 2007; **93**: 920-929. doi: 10.1529/biophysj.106.095927.

[28] **Flechsig H**, Popp D, Mikhailov AS. *In silico* investigation of conformational motions in superfamily 2 helicase proteins. *PLoS ONE* 2011; **6**(7): e21809. doi: 10.1371/journal.pone.0021809.

[29] **Gu M**, Rice CM. Three conformational snapshots of the hepatitis C virus NS3 helicase reveal a ratchet translocation mechanism. *Proc Natl Acad Sci USA* 2010; **107**: 521-528. doi: 10.1073/pnas.0913380107.

[30] **Appleby TC**, Anderson R, Fedorova O, Pyle AM, Wang R, Liu X, Brendza KM, Somoza JR. Visualizing ATP-dependent RNA translocation by the NS3 helicase from HCV. *J Mol Biol* 2011; **405**: 1139-1153. doi: 10.1016/j.jmb.2010.11.034.

[31] **Cheng W**, Arunajadai SG, Moffitt JR, Tinoco I, Bustamante C. Single-base pair unwinding and asynchronous RNA release by the hepatitis C virus NS3 helicase. *Science* 2011; **333**: 1746-1749. doi: 10.1126/science.1206023.

[32] **Zheng W**, Tekpinar M. Structure-based simulations of the translocation mechanism of the hepatitis C virus NS3 helicase along single-stranded nucleic acid. *Biophys J* 2012; **103**: 1343-1353. doi: 10.1016/j.bpj.2012.08.026.

[33] **Ando T**. Molecular machines directly observed by high-speed atomic force microscopy. *FEBS Lett* 2013; **587**: 997-1007. doi: 10.1016/j.febslet.2012.12.024.

[34] **Ando T**, Uchihashi T, Kodera N. High-speed AFM and applications to biomolecular systems. *Annu Rev Biophys* 2013; **42**: 393-414. doi: 10.1146/annurev-biophys-083012-130324.